%% file: grass19.tex
\documentclass[runningheads,a4paper]{llncs}

\usepackage{amssymb}
\usepackage{graphicx}
\usepackage{acronym}
\usepackage{url}
\usepackage{amsmath}
\usepackage{booktabs}
\usepackage{siunitx}
\usepackage{cleveref}
\usepackage{nth}
\usepackage{multirow}
\usepackage{amssymb,mathtools,amscd}
\usepackage{algorithm}
\usepackage{algorithmic}
\usepackage{subfigure}
\usepackage{longtable}
\usepackage{color}
\usepackage{url}
\usepackage{graphics}

\input{newcomma1}

\begin{document}
\mainmatter
\title{\vspace*{-0.99cm}\footnotesize{Accepted as a Conference Paper for ICONIP2017}\\
	\large A Grassmannian Approach to Zero-Shot Learning for Network Intrusion Detection}
\titlerunning{A Grassmannian Approach to Zero-Shot Learning } 

\urldef{\mailrivero}\path|rivero@dei.uc.pt|
\urldef{\mailbribeiro}\path|bribeiro@dei.uc.pt|
\urldef{\mailfleite}\path|fleite@mat.uc.pt|
\urldef{\mailnchenyx}\path|nchenyx@outlook.com|

\author{
	Jorge Rivero$^{\text{1}}$
	\and
	Bernardete Ribeiro$^{\text{1}**}$
	\and
	Ning Chen$^{\text{2}}$	
	\and
	F\'{a}tima Silva Leite$^{\text{3}}$
}


\institute{
  $^{1}$\footnote{Corresponding author}CISUC, Department of Informatics Engineering, University of Coimbra, Portugal\\
	$^{2}$College of Computer Science and Technology, Henan Polytechnic University, China\\
	$^{3}$Department of Mathematics\\and \\Institute of Systems and Robotics (ISR)\\
University of Coimbra, Portugal \\
	\mailrivero, \mailbribeiro, \mailnchenyx,  \mailfleite
}
\maketitle

\begin{abstract}
One of the main problems in Network Intrusion Detection comes from constant rise of new attacks, so that not enough labeled examples are available for the new classes of attacks. Traditional Machine Learning approaches hardly address such problem. This can be overcome with Zero-Shot Learning, a new approach in the field of Computer Vision, which can be described in two stages: the Attribute Learning and the Inference Stage. The goal of this paper is to propose a new Inference Stage algorithm for Network Intrusion Detection. In order to attain this objective, we firstly put forward an experimental setup for the evaluation of the Zero-Shot Learning in Network Intrusion Detection related tasks. Secondly, a decision tree based algorithm is applied to extract rules for generating the attributes in the AL stage. Finally, using a representation of a Zero-Shot Class as a point in the Grassmann manifold, an explicit formula for the shortest distance between points in that manifold can be used to compute the geodesic distance between the Zero-Shot Classes which represent the new attacks and the Known Classes corresponding to the attack categories. The experimental results in the datasets KDD Cup 99 and NSL-KDD show that our approach with Zero-Shot Learning successfully addresses the Network Intrusion Detection problem.
\end{abstract}

\section{Introduction}
The network intrusion detection systems (NIDS) are classified according to their detection type: (i) misuse detection, which monitors the activity with previous descriptions of known malicious behavior; (ii) anomaly detection, which  defines a profile for normal activity and looks for deviations; (iii) hybrid detection, resulting from a combination of the previous detection methods. The first two approaches have been widely studied and addressed by extensive academic research, yielding good results that rely on Machine Learning (ML) tools. However, the deployment of those systems in operational settings has been rather limited. For instance, one of the main challenges related to Network Intrusion Detection (NID) is the detection of new attacks. One possible way to face such incidents is to use outlier detection approaches. But these ML algorithms perform much better in matching similarities (misuse detection) than in finding activities that do not adjust to some predefined profile (anomaly detection). 

Zero-Shot Learning (ZSL) is a recent approach that has gained popularity to solve computer vision related tasks in which new classes may appear after the learning stage \cite{akata,lampert,eszsl}. Traditional ML cannot tackle these challenging scenarios.
ZSL uses an intermediate level called attributes. This level provides semantic information about the classes to classify. It is a way to identify new objects matching its descriptions in terms of attributes with concepts previously learned  \cite{eszsl}. Explained in a formal way, traditional classification learns, from labeled data, a mapping function $f:\, X \to Y$ from an input $x$, belonging to the space $X$, to an output $y$ in the class space $Y$. In ZSL there are no labeled samples for some classes in the space $Y$ and still a prediction is required \cite{akata}.

In this paper, we address the problem of detecting new attacks by a rather innovative use of a ZSL approach in two-stages: first, it uses the signatures of known attacks to learn the attributes; second, it makes inference of the new classes after some data transformation in the Grassmannian.  Our approach which can be considered a hybrid-based one has contributions  in both ZSL stages as it will be explained in the next sections.

The paper is organized as follows. In Section~\ref{sec:ZSL}, we present a brief overview of  the Zero--Shot Learning approach. Section~\ref{sec:grass} contains a short description of the Grassmannian, the manifold that will be used to represent the data. The experimental setup appears in Section~\ref{sec:data}.
In Section~\ref{sec:inid}, the Inference Network Intrusion Detection (INID) Algorithm for the application of ZSL is proposed. In Section~\ref{sec:results}, the results of INID evaluation on KDD Cup 99 and NSL-KDD datasets are discussed. Finally, in Section~\ref{sec:con}, we address the conclusions and future work.

\section{Zero-Shot Learning Background}
\label{sec:ZSL}
ZSL has simply two stages. The first  one is Attribute Learning (AL), where an intermediate layer that provides semantic information about the classes is learned. The goal is to capture knowledge from data.
The semantic information obtained from this stage is used in the second one, the Inference stage (IS), to classify instances among a new set of classes.
For modeling the relationships among features, attributes, and classes, different solutions have been proposed \cite{lampert,eszsl}.

The Attribute Learning (AL) stage in the context of NIDS was first proposed in  \cite{ALNID2016}. Therein,  different Machine Learning algorithms on two preprocessed datasets: (i) KDD Cup 99 and (ii) NSL-KDD  were evaluated. The best result was obtained from the evaluation of C45 decision tree algorithm with the classification accuracy of $99.54\%$.
The Attribute Learning for Network Intrusion Detection (ALNID) proposed therein is a rule-based algorithm which weights the attributes according to their entropy and frequency in the extracted rules.  The input to the algorithm ALNID is the set of instances $X = \{ A_i\}, i = 1\cdots m$. Each instance $A$ is composed of $d$ attributes $A = \{ a_1 ,\cdots,a_d \}$. The quantity of information (QI), the entropy (E) and the information gain (G) were computed for each $a_i  \in A$. During each iteration, the number of times that each $a_i$  attribute was evaluated by each rule of the set $R = \{ r_1 ,\cdots,r_j \}$ was recorded. With this frequency count (increasing by each time  an attribute is evaluated by the rule $r_j$) a new set of attributes $A' = \{ a_1 ',\cdots,a_d '\} $ is created. As a result, the algorithm returns the set of new valued instances $X = \{ A_i '\}, i = 1\cdots m$ \cite{ALNID2016}.

During the Inference Stage (IS), classes are inferred from the learned attributes. There are three approaches for this second stage \cite{eszsl}: K-Nearest Neighbour (K--NN), probabilistic frameworks \cite{lampert}, and energy function \cite{akata,eszsl}.
The K--NN inference consists in finding the closest test class signature to the predicted attribute signature found by mapping the input instance  into the attribute space in the previous stage.
In the cascaded  inference probabilistic framework proposed in \cite{lampert} the predicted attributes in the AL stage are combined to determine the most likely target class. The approach has two variants: (i) Directed Attribute Prediction (DAP), which learns a probabilistic classifier for each attribute during the AL stage. The classifiers are then used during the IS to infer new classes from their attributes signatures.
In another variant (ii) Indirected Attribute Prediction (IAP), the predictions from individual classifiers, one per each training class, are obtained. At test time, the predictions for all training  classes induce a labelling of the attribute layer from which a labelling over the test classes can be inferred~\cite{Lampert2009}.

In a recent approach \cite{akata,eszsl,crossmodal} based on energy function one considers at the training stage $z$ known classes (KC), which have a signature composed of $a$ attributes. Those signatures are represented by a matrix $ S \in [0,1]^{a\times z} $, where $[0,1]^{a\times z}$ denotes the set of matrices of order $a\times z$ whose entries belong to the real interval $[0,1]$.  The entries of $S$ represent the relationship between each attribute and the classes.
 The instances available at training stage are denoted by
a matrix $X \in \mathbb{R}^{d\times m} $,
where $d$ is the dimensionality of the data, and $m$ is the number of instances. The matrix $ Y \in \{  - 1,1\} ^{m\times z} $ is computed to indicate to which class each instance belongs, and during the training the following matrix $V \in \mathbb{R}^{d\times a}$, where $\gamma$ and $\lambda$ are hyper-parameters of the regularizer, is also computed\cite{eszsl}:
\begin{equation}
V = (XX^T  + \gamma I)^{ - 1} XYS^T(SS^T  + \lambda I)^{ - 1} .
\label{eq:eszsl}
\end{equation}
In IS, a new set of $z'$ classes is defined by their attributes signatures, $S' \in [0,1]^{a\times z'} $. Given a new instance $x$, the prediction is given by:
\begin{equation}
\mathop {\arg \max }\limits_i (x^T VS_i ').
\label{eq:3}
\end{equation}
\vspace*{-0.8 cm}
\section{Grassmann Manifold}
\label{sec:grass}
We introduce the manifold that will play the main role herein. The Grassmann manifold (or simply, the Grassmannian), hereafter denoted by $G_{k,m}$, is the set of all $k$-dimensional linear subspaces in $\mathbb{R}^m$ ($k\leq m$). This manifold, which doesn't have the geometry of an Euclidean space, has a matrix representation as:\\
\vspace*{-0.3 cm}
\beq
G_{k, m}  = \{ P \in \mathbb{R}^{m\times m} :P^2  = P, P^T  = P,\, rank(P) = k \}.
\label{eq:grassman}
\eeq
$G_{k,m}$ can be equipped with a metric inherited from the Euclidean metric on the vector space consisting of all $m\times m$ matrices.

There are other authors that do not identify a point in the Grassmann manifold with a subspace, but rather with an orthonormal frame that generates that subspace. In this case, points in the Grassmann are represented by rectangular $m\times k$ matrices whose columns are orthonormal. This is, for instance, the representation followed in \cite{Hospedales2015}.
Distances between two points $ P_1 $ and $ P_2 $ in  the Grassmannian are often defined using the notion of principal angle which can be calculated from the Singular Value Decomposition (SVD) of the matrix $P_1^TP_2$. A list of the most commom distance functions used in the literature, as well as a comparison among them, can be found, for instance, in \cite{grassmanndistance2}.

However, in the present article, besides the different representation of the Grassmann manifold as above, we  adopt  an alternative way, proposed in \cite{Batzies2015}, to compute distances in $G_{k,m}$.
More precisely, we use the following  closed formula for the geodesic distance between two points $P_1$ and $P_2$ in $G_{k,m}$, that depends on the two points only.
\vspace*{-0.3 cm}
\beq
d^2(P_1,P_2) =  - \frac{1}{4}trace \left(\log ^2 ((I - 2P_2)(I - 2P_1))\right) ,
\label{eq:function}
\eeq
where $\log$ is the principal logarithm of a matrix. Note that $d^2(P_1,P_2)\geq 0$.
This is due to the fact that, for $P_1, P_2\in G_{k,m}$,
the matrix  $(I - 2P_2)(I - 2P_1)$ is orthogonal and its $\log$ is skewsymmetric.
 So $\log ^2 ((I - 2P_2)(I - 2P_1))$ is symmetric
 with negative trace.
\section{Experimental Setup}
\label{sec:data}
\subsubsection{Network Intrusion Datasets.}
The KDD Cup 99 intrusion detection database\footnote{Available at \url{http://www.kdd.org/kddcup/ index.php}}, contains approximately $5$ million samples. This dataset
contains four different types of attacks: Denial Of Service (DOS), unauthorized access from a remote machine (R2L), U2R and probing.
Each instance represents a TCP/IP connection composed by $41$ features.
 A new dataset with $125,973$ selected records, NSL-KDD\footnote{\url{http://www.unb.ca/research/iscx/dataset/iscx-NSL-KDD-dataset.html}} solves the criticism of  data samples redundancy in  KDD Cup 99. Both datasets contain $23$ classes, where one class corresponds to normal traffic and the remainder $22$ classes represent attacks.
Some attacks have few records such as: \textit{spy} and \textit{perl} with just 2 and 3 instances respectively, while other classes such as: \textit{normal} and \textit{smurf} attack are represented by $97,277$ and $280,790$ respectively.
\begin{table*}[t]
	\centering
	\caption{Experimental NSL-KDD dataset setup for Zero-Shot Learning.}
	\label{table:Table4}
	\begin{tabular}{lcccr}
		\hline
		\multicolumn{1}{c}{\textbf{\begin{tabular}[c]{@{}c@{}}Category of\\  Attacks\end{tabular}}} & \textbf{\begin{tabular}[c]{@{}c@{}}Known\\  Classes\end{tabular}} & \textbf{\begin{tabular}[c]{@{}c@{}}Instances per \\ Known Classes\end{tabular}} & \textbf{\begin{tabular}[c]{@{}c@{}}Zero-Shot \\ Classes\end{tabular}} & \multicolumn{1}{c}{\textbf{\begin{tabular}[c]{@{}c@{}}Instances per \\ Zero-Shot Classes\end{tabular}}} \\ \hline
		\multirow{4}{*}{DoS}                                                                        & smurf                                                             & 2646                                                                            & \multirow{2}{*}{teardrop}                                             & \multirow{2}{*}{892}                                                                                    \\
		& neptune                                                           & 41,214                                                                          &                                                                       &                                                                                                         \\
		& back                                                              & 956                                                                             & \multirow{2}{*}{land}                                                 & \multirow{2}{*}{19}                                                                                     \\
		& pod                                                               & 201                                                                             &                                                                       &                                                                                                         \\ \hline
		Normal                                                                                      & normal                                                            & 67,343                                                                          & \multicolumn{1}{l}{-}                                                 & \multicolumn{1}{l}{-}                                                                                   \\ \hline
		Probe                                                                                       & satan                                                             & 3632                                                                            & ipsweep                                                               & 3599                                                                                                    \\
		& portsweep                                                         & 2931                                                                            & nmap                                                                  & 1493                                                                                                    \\ \hline
		\multirow{6}{*}{R2L}                                                                        & warezclient                                                       & 890                                                                             & \multirow{3}{*}{guess\_passwd}                                        & \multirow{3}{*}{53}                                                                                     \\
		& warezmaster                                                       & 20                                                                              &                                                                       &                                                                                                         \\
		& ftp\_write                                                        & 8                                                                               &                                                                       &                                                                                                         \\
		& multihop                                                          & 8                                                                               & \multirow{3}{*}{imap}                                                 & \multirow{3}{*}{648}                                                                                    \\
		& phf                                                               & 4                                                                               &                                                                       &                                                                                                         \\
		& spy                                                               & 2                                                                               &                                                                       &                                                                                                         \\ \hline
		U2R                                                                                         & buffer\_overflow                                                  & 30                                                                              & rootkit                                                               & 10                                                                                                      \\
		& loadmodule                                                        & 9                                                                               & perl                                                                  & 3                                                                                                       \\ \hline
	\end{tabular}
\end{table*}
Table~\ref{table:Table4} presents the attack categories and instances and  for NSL-KDD.
\subsubsection{Data Preprocessing.}
In \cite{ALNID2016} a data setup was proposed for the AL stage of ZSL in NIDS. Similarly, the preprocessing ifor  NSL-KDD\footnote{Due to space limitations we present only the NSL-KDD setup for Zero-Shot Learning.} is:
\begin{enumerate}
		\item Taking into account \cite{ALNID2016}, the selected attributes from the original $41$ were: $1$, $2$, $5$, $6$, $9$, $23$, $24$, $29$, $32$, $33$, $34$ and $36$.	
	\item For each category of network attack, the instances which represent two classes of attacks were removed -- the Zero Shot Classes (ZSC) 
	(see Table~\ref{table:Table4}\footnote{Note that by adding the number of KC (15, 2nd column) and the number ZSC (8, 4th column) we obtain the total number of 23 classes for this dataset.})
	\item The original attack labels were replaced by their corresponding category (e.g. labels such as: \textit{smurf}, \textit{neptune}, \textit{back} and \textit{pod} were replaced by their category label: \textit{DoS}). The categories are the Known Classes (KC).	
	\item The datasets were split into different files by categories. %
\end{enumerate}
\subsubsection{Data Transformation.}
\label{sec:datatrans}
In this section we explain how to associate data to points in $G_{k,m}$.
The data transformation is described in the following steps:
\begin{enumerate}
\item A set of data is first represented as a rectangular matrix $ X_{m\times d}$, where $m$ is the number of instances and $d$ the number of attributes.
\item The matrix $ X_{m\times d}$ is decomposed using the Singular Value Decomposition
\beq X_{m\times d} = U_{m\times m} \, \Sigma_{m\times d} \, V_{d\times d}^T, \label{eq:svd} \eeq
 the matrices $U$ and $V$ are orthogonal ($UU^T=I_m$, $VV^T=I_d$) and   $\Sigma$ is a quasi-diagonal matrix containing the singular values $\sigma _1, \cdots , \sigma_d$ of $X$, in non-increasing order, along the main diagonal. Since $XX^T=U(\Sigma \Sigma^T ) U^T$, the columns of $U$ are the eigenvectors associated to the eigenvalues $\lambda _i$ of $XX^T$, which are the square of the singular values of $X$ and are, by convention,  also descendent sorted ($\lambda _1  \ge \lambda _2  \ge ... \ge \lambda _m \ge 0$). The columns of the matrix $U_{m\times m}$ are called the eigenvectors of the SVD decomposition.
\item  Since the first columns of $U$ are the most significant directions, we define a threshold $
0 < \alpha  \le 100$ that selects the most important eigenvectors of $U$ and with them we form the submatrix $S_{m\times k} $, whose columns form an  orthonormal set of $k$ vectors in $\mathbb{R}^m$, i.e, $S^TS=I_k$, and $k$ must satisfy the condition
\beq {{\sum\limits_{i = 0}^k {\lambda _i } } \over {\sum\limits_{i = 0}^m {\lambda _i } }}*100 \ge \alpha . \label{eq:pca} \eeq
\item From the previous matrix $S_{m\times k}$, we compute a square matrix  $P_{m\times m}=SS^T $, which can easily be proven to belong to the Grassmann manifold
$ G_{k, m}$.
This matrix $P_{m\times m}$ gives a representation of the data in the Grassmannian.
\end{enumerate}
Having the data represented in the Grassmannian, the formula (\ref{eq:function}) for the geodesic distance between points in that manifold can be used to compute the shortest distance between the zero-shot classes and the known classes.
\begin{figure*}[t!]
	\centering
    \includegraphics[keepaspectratio, width = 12cm]{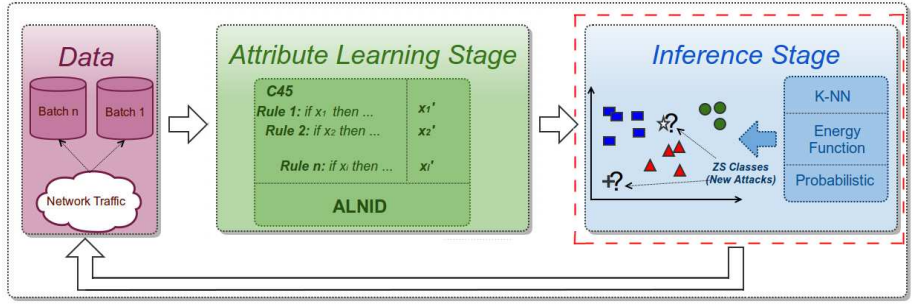}
	\caption{Zero-Shot Learning framework\cite{ALNID2016}. Left:  network traffic datasets; Middle: Attribute learning Stage; Right: Inference Stage.}
	\label{fig:zero-shot}
\end{figure*}
\begin{algorithm}
\begin{algorithmic}[1]		
		\REQUIRE   Z  ,   K  ,   cutoff\_percent  		
		\STATE   batch\_start, batch\_stop = 0
		\STATE   distances\_list = [ ]
		\STATE   batches = K.rows / Z.rows
		\FOR{  batch   in   batches  }
		\STATE   batch\_stop = Z.rows + batch\_start
		\STATE   data\_batch = K[batch\_start:batch\_stop]  			
		\STATE   U\_Z, s\_Z, V\_Z = SVD(Z)
		\STATE   U\_K\_batch,s\_K\_batch, V\_K\_batch=SVD(data\_batch)  		
		\STATE   columns = SelectEigenValues(U\_Z, s\_Z, U\_K\_batch, s\_K\_batch, cutoff\_percent)
		\STATE   $P_2$=U\_Z[:,0:columns]*U\_Z[:,0:columns].T
		\STATE   $P_1$ = U\_K\_batch[:,0:columns] * U\_K\_batch[:,0:columns].T
		\STATE   distance =$sqrt(-1/4 * trace(\log ^2 ((I - 2P_2)(I - 2P_1)))) $
		\STATE   distances\_list.append(distance)
		\STATE   batch\_start = batch\_stop + 1
		\ENDFOR
		\RETURN   mean(distances\_list)  	
	\end{algorithmic} \caption{INID: Inference for Network Intrusion
	Detection}\label{alg:inid} \end{algorithm}
\section{Proposed Inference Algorithm Based on Grassmannian}
\label{sec:inid}
In this section we propose an Inference Network Intrusion Detection (INID) Algorithm for the second stage of the Zero Shot Learning approach. In Figure~\ref{fig:zero-shot} the overall tasks are depicted including the two-stages identified as Attribute Learning  and Inference Learning.
The algorithm~\ref{alg:inid} INID requires as \textbf{inputs} the results from the first stage,  i.e., the outputs of the ALNID algorithm \cite{ALNID2016} which tackles the attribute learning problem, and \textbf{outputs} the mean distances using the geodesic distance between two points in the Grassmannian given by formula (\ref{eq:function}).
Algorithm~\ref{alg:inid} INID begins with the data transformation required to map the data into a Grassmannian. It requires two sets of learned attributes from Algorithm ALNID which takes  both preprocessed datasets for learning the attributes needed for the IS stage. These are the learned attributes of ZSC (Algorithm~\ref{alg:inid}, $Z$) and the learned attributes of KC (Algorithm~\ref{alg:inid}, $K$). Then, these matrices are processed in different batches with $X_{m\times d}$ matrix (Algorithm~\ref{alg:inid}, lines: 1--6). In the sequel, for each batch, the factorization step is performed by the Singular Value Decomposition (Algorithm~\ref{alg:inid}, lines: 7--8). Hereafter, the sub-matrix $S_{m\times k}$ is computed by selecting the first $k$ columns (according to (\ref{eq:pca}), Algorithm~\ref{alg:inid}, line: 9) of $U$, where $\alpha $ is the cut-off$\_$percent. In the following step,  the matrices $P_1$ and $P_2=SS^T$, which live in the Grassmannian defined in~(\ref{eq:grassman}), are computed (Algorithm~\ref{alg:inid}, lines: 10--11).
After this data transformation, the distance between $P_1$ and $P_2$ is computed (formula (\ref{eq:function}), Algorithm~\ref{alg:inid}, line: 12). The algorithm returns the mean of the distances between $Z$ and all batches computed from $K$ (Algorithm~\ref{alg:inid}, line: 16).
\begin{table}[]
	\centering
	\caption{Evaluated distance functions}
	\label{table:distancefunctions}
	\begin{tabular}{ll}
		\hline
		\textbf{Distance Functions} & \multicolumn{1}{c}{\textbf{Formula}}                                                                                                      \\ \hline
		Frobenius Distance  & $ d^2_F (X_1,X_2) = \parallel X_1 - X_2 \parallel _F=trace((X_1 - X_2)^T(X_1 - X_2))$\\
		Grassmannian Distance                     & $
		d^2(P_1,P_2) =  - \frac{1}{4}trace \left(\log ^2 ((I - 2P_2)(I - 2P_1))\right) \label{eq:fatima}$ \\
		 \hline
	\end{tabular}
\end{table}
\begin{table}[p]
	\centering
	\caption{Grassmannian computed distances mean on KDD Cup 99 dataset.}
	\begin{tabular}{lcccccr}
		\hline
\textbf{Zero Shot Class  (ZSC)} & \textbf{Norma}l & \textbf{DoS} & \textbf{Probe} & \textbf{R2L} &	\textbf{U2R} & \\	\hline
teardrop (DoS)	&	3.0608	&	\textbf{2.5190}	&	4.0247	&	2.5231	&	2.5588	\\
land (DoS)	&	2.7603	&	\textbf{2.4949}	&	2.6493	&	2.5601	&	2.5600	\\
ipsweep (Probe)	&	0.7081	&	0.0587	&	\textbf{0.0539}	&	2.4769	&	2.5298	\\
nmap (Probe)	&	2.8888	&	2.5347	&	\textbf{2.5012}	&	3.0173	&	2.8697	\\
guess\_passwd (R2L)	&	0.5588	&	0.1637	&	0.6675	&	\textbf{0.0249}	&	2.5327	\\
imap (R2L)	&	2.6924	&	2.5770	&	2.6014	&	\textbf{2.5160}	&	3.5102	\\
rootkit (U2R)	&	2.6239	&	2.5033	&	3.4797	&	2.5479	&	\textbf{2.4976}	\\
perl (U2R)	&	0.1895	&	\textbf{\color{red}{\fbox{0.0090}}}	&	0.0717	&	0.5869	&	0.0581	\\
		\hline
	\end{tabular}
	\label{table:Table2}
\end{table}
\begin{table}[p]
	\centering
	\caption{Grassmannian computed distances mean on NSL-KDD dataset.}
	\begin{tabular}{lcccccr}
		\hline
\textbf{Zero Shot Class  (ZSC)} & \textbf{Norma}l & \textbf{DoS} & \textbf{Probe} & \textbf{R2L} &	\textbf{U2R} & \\	\hline
teardrop (DoS)	&	4.4785	&	\textbf{2.5300}	&	4.7464	&	4.8005	&	4.6889	\\
land (DoS)	&	3.2897	&	\textbf{2.6301}	&	3.9121	&	3.6518	&	4.4227	\\
ipsweep (Probe)	&	2.5609	&	2.4950	&	\textbf{0.0605}	&	2.4812	&	2.5167	\\
nmap (Probe)	&	4.6949	&	4.8336	&	\textbf{2.5280}	&	4.7910	&	4.7170	\\
guess\_passwd (R2L)	&	1.5401	&	2.4461	&	2.5099	&	\textbf{0.0655}	&	2.5195	\\
imap (R2L)	&	3.1187	&	3.4144	&	3.5278	&	\textbf{2.5430}	&	2.6641	\\
rootkit (U2R)	&	3.0829	&	3.2873	&	3.4814	&	3.3095	&	\textbf{2.6847}	\\
perl (U2R)	&	0.8621	&	0.8786	&	1.6779	&	\textbf{\color{red}{\fbox{0.3715}}}	&	1.3230\\ \hline
\end{tabular}
	\label{table:Table5}
\end{table}
\begin{table}[p]
\centering
\caption{Frobenius computed distances mean on KDD Cup 99 dataset.}
	\label{table:Frobenius1}
\begin{tabular}{lcccccr}
	\hline
	\textbf{Zero Shot Class  (ZSC)} & \textbf{Norma}l & \textbf{DoS} & \textbf{Probe} & \textbf{R2L} &	\textbf{U2R} & \\	\hline
teardrop (D)	&	0.1205	&	\textbf{0.0468}	&	0.8548	&	1.3097	&	1.0172	\\
land (D)	&	1.4359	&	\textbf{0.4416}	&	1.6868	&	1.7090	&	2.4949	\\
ipsweep (P)	&	2.9133	&	\textbf{\color{red}{\fbox{0.1449}}}	&	0.2424	&	2.7623	&	1.7281	\\
nmap (P)	&	1.7207	&	0.4452	&	\textbf{0.0397}	&	2.7267	&	2.0296	\\
guess\_passwd (R)	&	1.6629	&	1.1183	&	\textbf{\color{red}{\fbox{0.4993}}} &	1.7366	&	2.0681	\\
imap (R)	&	\textbf{\color{red}{\fbox{0.7023}}}	&	0.7212	&	0.7928	&	1.7694	&	2.3412	\\
rootkit (U)	&	2.6602	&	0.3621	&	\textbf{\color{red}{\fbox{0.0367}}}	&	2.8921	&	2.5664	\\
perl (U)	&	1.4898	&	\textbf{\color{red}{\fbox{0.6870}}}	&	1.1751	&	2.9581	&	1.9001 \\	\hline
\end{tabular}
\end{table}
\begin{table}[p]
	\centering
	\caption{Frobenius computed distances mean on NSL-KDD dataset.}
	\begin{tabular}{lcccccr}
		\hline
		\textbf{Zero Shot Class (ZSC)} & \textbf{Norma}l & \textbf{DoS} & \textbf{Probe} & \textbf{R2L} &	\textbf{U2R} & \\ \hline
teardrop (D)	&	\textbf{0.3541}	&	1.0402	&	0.7990	&	2.2335	&	1.8625	\\
land (D)	&	1.6816	&	\textbf{0.3378}	&	1.8476	&	1.7114	&	1.7814	\\
ipsweep (P)	&	2.1601	&	\textbf{\color{red}{\fbox{0.2042}}} &	0.6908	&	2.1157	&	2.9321	\\
nmap (P)	&	1.9747	&	1.5818	&	\textbf{0.2457}	&	2.0880	&	2.3004	\\
guess\_passwd (R)	&	\textbf{\color{red}{\fbox{0.4146}}}	&	1.8328	&	1.6318	&	2.7305	&	1.8766	\\
imap (R)	&	0.5065	&\textbf{\color{red}{\fbox{0.2485}}}	&	0.8849	&	1.2402	&	1.6496	\\
rootkit (U)	&	\textbf{\color{red}{\fbox{0.2427}}} &	0.7338	&	1.2477	&	1.4055	&	2.5617	\\
perl (U)	&	\textbf{\color{red}{\fbox{1.2974}}}	&	2.2123	&	2.6119	&	2.8423	&	2.8073\\	\hline
\end{tabular}
	\label{table:Frobenius2}
\end{table}
\section{Results and Discussion}
\label{sec:results}
The evaluation of our proposal was done in different batches. Each batch size was determined by the number of instances of the ZSC class.
In order to validate the results, the means of the distances among all the batches of each category and the ZSC were computed.
In Table~\ref{table:distancefunctions} we highlight the formula for the Frobenius distance, commonly used in ML (see, for instance, \cite{eszsl,morell}, and the new formula for the Grassmannian distance that is used in this paper after data transformation (see Section \ref{sec:datatrans}).
 In Table~\ref{table:Table2}, the evaluation results of our approach in the KDD Cup 99 dataset are presented. It is observed that for each ZSC the shortest distance computed corresponds to its respective category of attack. We emphasize  that this occurs for almost all the ZSC considered: teardrop (DoS), land(DoS), issweep (Probe), nmap (Probe), guess\_passwd (R2L), imap (R2L), rootkit (U2R) (in bold). Since only three instances of the ZSC \textit{perl} were evaluated, the shortest distance was wrongly 'assigned' to the attack category DoS (in red). Likewise, in the NSL-KDD dataset (see Table~\ref{table:Table5}) only the ZSC \textit{perl} is wrongly 'assigned' to class R2L (in red). A close look at each row Table~\ref{table:Table2} reveals that the second shortest distance for most of the ZSC in KDD Cup 99 was in DoS category showing its redundancy. For comparison Tables~\ref{table:Frobenius1} and~\ref{table:Frobenius2} illustrate the misplacements of ZSC (in red) relatively to the ground truth when Frobenius distance is used.

For the prediction models, two K-NN algorithms were used. The first one, with the geodesic distance in the Grassmannian proposed in \cite{Batzies2015}, and the second one, with the Frobenius distance function both illustrated in Table~\ref{table:predictionresults} with the performance results. Both datasets were respectively split into $70$\% for training
and the remaining $30$\% for test. 
For each evaluation the distance between each new data instance and the instances previously stored is computed.
\vspace*{-0.5 cm}
\begin{table}[htb!]
	\centering
	\caption{K-NN Prediction Model Performance (K was set to $5$)}
	\label{table:predictionresults}
	\begin{tabular}{lccccr}
		\hline
		\textbf{Metrics}        & \textbf{ Grassmannian Distance} && \textbf{Frobenius Distance}  \\ \hline
		Classification Accuracy &   \textbf{90.61}\%              & &           82.93\%          &                    &                    \\
		Logarithmic Loss        &    0.228             &&          \textbf{0.122}              &                  &                    \\
		AUC    &  \textbf{86.1}\%               &&        68.5\%              &                              &                    \\ \hline
	\end{tabular}
\end{table}
The Table~\ref{table:predictionresults} shows the performance metrics evaluated for each prediction model. Firstly with a classification accuracy of $90.61\%$, our proposal shows better performance than the one based on Frobenius distance, while showing a small Logarithmic Loss. Finally, the AUC metric shows very good results with a value of $86.1\%$.

The Grassmannian approach validates the AL algorithm proposed in \cite{ALNID2016} and the data transformation into the Grassmann manifold. Moreover, our results reveal the potential of the mathematical formula (\ref{eq:function}) in~\cite{Batzies2015} that computes the geodesic distance in the Grassmann manifold, with better results than other approaches. This might be of particular importance to solve other problems in ML that require computing distances among data points.
\section{Conclusions}
\label{sec:con}
The need for detecting new attacks on traffic networks, together with the competence of the ZSL approach to classify new classes without any example during training, motivated its application to NIDS. This study takes into account our previous results of the ALNID algorithm proposed in \cite{ALNID2016} for the AL stage which provided a significant improvement in the attributes representation. The good class separability found by the learned attributes lead us to look for an instance-based inference stage. We then resorted to a simplified and non-overlapping representation of data points that lie along a low-dimensional manifold embedded in a high-dimensional attributes space. Therefore, we transformed the learned attributes and represented them on the Grassmannian. We proposed the INID algorithm for the IS stage of the ZSL approach using KDD Cup 99 and NSL-KDD datasets. The algorithm computes the Grassmannian distances between the known attack KC and the selected ZSC both represented as points in the Grassmann manifold. The  prediction results by K-NN show that our method excels in performance as compared to Frobenius distance computed in the same predictor. Further work will explore the manifolds' representation.
\vspace*{-0.3 cm}
\section*{Acknowledgement}  Erasmus Mundus Action 2 is acknowledged for partial funding of the first author.

\noindent SASSI Project (33/SI/2015\&DT) is gratefully acknowledged for partial financial support.
\bibliographystyle{unsrt}
\bibliography{zsl1}
\end{document}

%% file: newcomma1.tex
\newcommand{\bi}{\begin{itemize}}\newcommand{\ei}{\end{itemize}}
\newcommand{\be}{\begin{enumerate}}\newcommand{\ee}{\end{enumerate}}
\newcommand{\bs}{\begin{slide}}\newcommand{\es}{\end{slide}}
\newcommand{\bc}{\begin{center}}\newcommand{\ec}{\end{center}}
\newcommand{\beq}{\begin{equation}}\newcommand{\eeq}{\end{equation}}
\newcommand{\beqn}{\begin{eqnarray}}\newcommand{\eeqn}{\end{eqnarray}}
\newcommand{\btab}{\begin{tabular}}\newcommand{\etab}{\end{tabular}}

